\documentstyle[sprocl]{article}

\begin{document}


\title{FERMION-FERMION AND BOSON-BOSON AMPLITUDES:
SURPRISING SIMILARITIES\,$^{\,\ast}$}

\author{VALERI V. DVOEGLAZOV\,$^{\,\dagger}$}

\address {
Escuela de F\'{\i}sica, Universidad Aut\'onoma de Zacatecas \\
Antonio Doval\'{\i} Jaime\, s/n, Zacatecas 98068, ZAC., M\'exico\\
Internet address:  VALERI@CANTERA.REDUAZ.MX
}


\maketitle\abstracts{Amplitudes for boson-boson and fermion-boson
interactions are calculated in the second order of perturbation theory in
the Lo\-ba\-chev\-sky space.
An essential ingredient of the used model is the Weinberg's
$2(2j+1)$ component formalism for describing a particle
of spin $j$, recently developed substantially.
The boson-boson amplitude is then compared
with the two-fermion amplitude obtained  by Skachkov long ago
on the ground of the hamiltonian formulation of quantum field
theory on the mass hyperboloid, $p_0^2 -{\vec p}^{\,2}=M^2$, proposed
by Kadyshevsky. The pa\-ra\-met\-ri\-za\-tion of the amplitudes by means
of the momentum transfer in the Lo\-ba\-chev\-sky space leads to
same spin structures in the expressions of $T$ matrices
for the fermion and the boson cases. However,  certain
differences are found. Possible physical applications are discussed.}


\noindent
The scattering amplitude for the two-fermion interaction had been
obtained in the Lobachevsky space in the second order of perturbation
theory  long ago, ref.~[1a,Eq.(31)]:
\begin{eqnarray}\label{eq:TF}
\lefteqn{T^{(2)}_V ({\vec q} (-) {\vec p}, {\vec p}) =
-g_v^2 \frac{4m^2}{\mu^2 +4 {\vec \kappa}^{\,2}} -
4g_v^2\frac{({\vec \sigma}_1 {\vec \kappa})({\vec \sigma}_2
{\vec \kappa}) - ({\vec \sigma}_1 {\vec \sigma}_2)
{\vec \kappa}^2}{\mu^2 +4{\vec \kappa}^{\,2}} -\nonumber}\\
&-& {8g_v^2 p_0 \kappa_0 \over m^2}\,
\frac{i{\vec \sigma}_1 [{\vec p} \times {\vec \kappa} ] +i{\vec \sigma}_2
[{\vec p} \times {\vec \kappa} ]}{\mu^2 +4 {\vec \kappa}^{\,2}} -
{8g_v^2 \over m^2}\,\frac{p_0^2 \kappa_0^2 +2p_0 \kappa_0 ({\vec p}
\cdot {\vec \kappa}) - m^4}{\mu^2 +4{\vec \kappa}^{\,2}} -\nonumber\\
&-& \frac{8g_v^2}{m^2}\,\frac{({\vec \sigma}_1 {\vec p})
({\vec \sigma}_1 {\vec \kappa}) ({\vec \sigma}_2 {\vec p})
({\vec \sigma}_2 {\vec \kappa})}{\mu^2 +4{\vec \kappa}^{\,2}}\quad,
\end{eqnarray}
$g_v$ is the coupling constant.
The treatment  is based on the use of the formalism
of separation of Wigner rotations and parametrization
of currents by means of the Pauli-Lyuban'sky vector, developed in
the sixties.~\cite{Shirokov} The quantities
$$\kappa_0 = \sqrt{\frac{m(\Delta_0 +m)}{2}}\quad,\quad
{\vec \kappa} = {\vec n}_\Delta \sqrt{\frac{m(\Delta_0 -m)}{2}}$$
are the components of  the 4-vector of momentum ``half-transfer".
This concept is closely connected with a notion of the half-velocity
of a particle.
The 4-vector $\Delta_{\mu}$:
\begin{eqnarray}
{\vec \Delta} &=& \Lambda^{-1}_{{\vec p}} {\vec q}
= {\vec q} (-) {\vec p} = {\vec q}
-\frac{{\vec p}}{m} (q_0 - \frac{{\vec q}\cdot
{\vec p}}{p_0 +m})\quad,\\
\Delta_0 &=& (\Lambda^{-1}_{p} q)_0 = (q_0 p_0
-{\vec q}\cdot{\vec p})/m = \sqrt{m^2\,+ \,{\vec \Delta}^2}
\end{eqnarray}
could be regarded as the momentum transfer vector in
the Lobachevsky space.
The amplitude (\ref{eq:TF}) has been  successfully applied for describing
bound states of two fermions in the framework of the Kadyshevsky version
of the quasipotential approach.  Moreover,  the use of the
Shapiro~\cite{Shapiro} technique of expansion in the plane-waves on
hyperboloid  and  of the supplementary series of  unitary representations
of the Lorentz group led Prof. Skachkov to the very interest model of the
quark confinement.~\cite{Skachkov1}

In order to obtain the 4-vector current for the interaction
of a Joos-Weinberg $j=1$ boson with the external 4-potential field
one can use the known formulas of refs.~\cite{Skachkov,Shirokov},
which are valid for any spin:
\begin{equation}
{\cal U}^\sigma({\vec p}) =
{S}_{{\vec p}} \,{\cal U}^\sigma({\vec 0})\quad, \quad {S}_{{\vec
p}}^{-1} {S}_{{\vec q}} = {S}_{{\vec q}(-){\vec p}}\cdot I\otimes
D^{J}\left \{ V^{-1}(\Lambda_p, q)\right \}\quad,
\end{equation}
\begin{equation}
W_\mu({\vec p})\cdot D^J\left \{ V^{-1}(\Lambda_{p}, q)\right \}
= D^J\left \{ V^{-1}(\Lambda_{p}, q)\right \}
\cdot\left [ W_\mu({\vec q})
-\frac{p_\mu+q_\mu}{M(\Delta_0+M)}p^\nu W_\nu ({\vec q})\right ],
\end{equation}
\begin{equation}
q^\mu W_\mu ({\vec p})\cdot D^J\left \{ V^{-1}(\Lambda_{p}, q)\right \} =
-D^J\left \{ V^{-1}(\Lambda_{p}, q)\right \}\cdot p^\mu W_\mu ({\vec q})
\quad.
\end{equation}
$W_\mu$ is the Pauli-Lyuban'sky 4-vector of relativistic spin;
the matrix \linebreak $D^J \left \{ V^{-1} (\Lambda_{p}, q)\right \}$
is the matrix of the Wigner rotation for spin $j$.

Thus, we come to the 4-current vector for a $j=1$ Joos-Weinberg boson:
\begin{eqnarray}
j_{\mu}^{\sigma_{p}\nu_{p}}({\vec p}, {\vec q}) &=&
j_{\mu \,(S)}^{\sigma_{p}\nu_{p}}({\vec p}, {\vec q}) +
j_{\mu \,(V)}^{\sigma_{p}\nu_{p}}({\vec p}, {\vec q}) +
j_{\mu \,(T)}^{\sigma_{p}\nu_{p}}({\vec p}, {\vec q})\quad,\\
j_{\mu \,(S)}^{\sigma_{p}\nu_{p}}({\vec p}, {\vec q}) \,&=&\,
-\,g_S \xi^\dagger_{\sigma_p} \left \{  (p+q)_\mu \left (
1+ \frac{({\vec J}\cdot {\vec \Delta})^2}{M (\Delta_0 + M)} \right )\right
\} \xi_{\nu_p}\quad,\label{curs}\\
j_{\mu \,(V)}^{\sigma_{p}\nu_{p}}({\vec
p}, {\vec q}) \,&=&\, -\,g_V \xi^\dagger_{\sigma_p} \left \{ (p+q)_{\mu}+
{1\over M}W_{\mu}({\vec p})({\vec J}\cdot{\vec \Delta})-
{1\over M}({\vec J}\cdot{\vec \Delta}) W_{\mu}({\vec p})\right \}
\xi_{\nu_p}\nonumber\\
&&\label{cur}\\
j_{\mu \,(T)}^{\sigma_{p}\nu_{p}}({\vec
p}, {\vec q}) \,&=&\, -\, g_T \xi_{\sigma_p}^\dagger \left \{ - (p+q)_\mu
\frac{({\vec J}\cdot {\vec \Delta})^2}{M (\Delta_0 + M)}+
\right.\label{curt}\\
&& \left. \qquad\qquad\qquad + {1\over M} W_{\mu}({\vec p})({\vec
J}\cdot{\vec \Delta})- {1\over M}({\vec J}\cdot{\vec \Delta})
W_{\mu}({\vec p})\right \} \xi_{\nu_p}\quad.\nonumber
\end{eqnarray}
Let us note an interesting feature. The 6-spinors
${\cal U} ({\vec p})$ and ${\cal V} ({\vec p})$ in the old Weinberg
formulation do not form a complete set:
\begin{equation}
{1\over M} \left \{{\cal U} ({\vec p}) \overline {\cal U} ({\vec p})
+{\cal V} ({\vec p}) \overline {\cal V} ({\vec p}) \right \} \,=\,
\pmatrix{I & {S}_{{\vec p}}\otimes {S}_{{\vec p}} \cr
{S}^{-1}_{{\vec p}}\otimes {S}^{-1}_{{\vec p}} &
I\cr}\quad.
\end{equation}
But, if  regard $\tilde{{\cal V}} ({\vec p})=
\gamma_5 {\cal V} ({\vec p})$ one can obtain
the complete set. Fortunately,
$\overline{\tilde {\cal V}} ({\vec 0}) {\cal U} ({\vec 0}) = 0$\, ,
what  permits us to keep the parametrization
\begin{equation}\label{chalf}
j^\mu_{\sigma\sigma^\prime} ({\vec p}, {\vec q}) =
\sum_{\sigma_p=-1/2}^{1/2} j^\mu_{\sigma\sigma_p} ({\vec q} (-) {\vec p},
{\vec p}) \,\,D^{1/2}_{\sigma_p \sigma^\prime}
\left \{ V^{-1} (\Lambda_p, q)\right \}\quad.
\end{equation}
As a matter of fact, this definition of negative-energy spinors follows from
the explicit construct of the theory of the Bargmann-Wightman-Wigner type,
which has been proposed by Ahluwalia,~\cite{DVA} and it presents itself
a generalization (and  a  correction) of the Joos-Weinberg model.

As a result we obtain the amplitude for   interaction of two
$j=1$ Joos-Weinberg particles, mediated by the vector potential:
\begin{eqnarray}\label{212}
\lefteqn{\hat T^{(2)} ({\vec q}(-){\vec p}, {\vec p})
\,=\, g^2 \left\{ \frac{\left [p_0
(\Delta_0 +M) + ({\vec p}\cdot {\vec \Delta})\right ]^2
-M^3 (\Delta_0+M)}{M^3 (\Delta_0 -M)}+\right.}\nonumber\\
&+&\left.\frac{i ({\vec J}_1+{\vec J}_2)\cdot\left [{\vec p}
\times{\vec \Delta}\right ]}
{\Delta_0-M}\left [ \frac{p_0 (\Delta_0 +M)+{\vec p}\cdot
{\vec \Delta}}{M^3} \right ] + \right.\nonumber\\
&+& \left.\frac{({\vec J}_1\cdot {\vec \Delta})({\vec
J}_2\cdot {\vec \Delta})-({\vec J}_1\cdot{\vec J}_2) {\vec \Delta}^2}{2M
(\Delta_0-M)}-\frac{1}{M^3}\frac{{\vec J}_1\cdot\left [{\vec p}
\times{\vec \Delta}\right ] \,\,{\vec J}_2\cdot \left [{\vec
p} \times{\vec \Delta}\right ]}{\Delta_0-M}\right\}.
\end{eqnarray}
We have assumed $g_S = g_V = g_T$ above, what is motivated
by group-theoretical reasons. The expression (\ref{212}) reveals  the
advantages of the $2(2j+1)$- formalism, since
it looks like  the amplitude for interaction of two spinor particles
within the substitutions
$1/\left [ 2M(\Delta_0 - M)\right ] \Rightarrow 1/{\vec \Delta}^2$ and
${\vec J}\Rightarrow {\vec \sigma}$.
Calculations hint that many analytical results produced for
a Dirac fermion could be applicable for describing a $2(2j+1)$
particle. Nevertheless, it is required adequate explanation
of the obtained difference. The  reader could note:
its origin lies at the kinematical
level.  Free-space (without interaction) Joos-Weinberg equations admit
acausal tachyonic solutions, {\it e.g.}, ref.~\cite{DVA0}. ``Interaction
introduced in the massive Weinberg equations will couple to both the
causal and acausal solutions and thus cannot give physically
reasonable results". However, let us not forget that we have used
the Tucker-Hammer approach, indeed, that does not possess
tachyonic solutions.

In the straightforward manner we also obtain  amplitudes for interactions:
1) \, spin-0 boson and spin-1/2 fermion; \,
2) \, spin-0 boson and spin-1 Joos-Weinberg boson; \,
3) \, spin-1/2 fermion and spin-1 Joos-Weinberg boson. \,
Like the previous ones the Wigner rotations are separated out  and all
spin indices are ``re-setted" on the momentum ${\vec p}$.
The detailed discussion of the presented topics could be found
in refs.~\cite{DVO1,DVO2,DVO3}.

\section*{Acknowledgments}

I  appreciate very much discussions with Prof.
D. V.  Ahluwalia and  Prof. A.~F. Pashkov.
I should thank  Prof. N. B. Skachkov for his help in analyzing
several topics. I am grateful to Zacatecas University for
professorship.

\medskip

\footnotesize{
$^{\,\ast}$ Reported at the IV Wigner Symposium,  Guadalajara,   M\'exico
(Aug. 7-11, 1995). A certain part of the work has been presented at the
Workshop ``Mathematical problems of statistical mechanics and quantum
field theory", Kuibyshev, USSR (May 19-21, 1987).\quad
$^{\,\dagger}$ On leave of
absence from {\it Dept. Theor. \& Nucl. Phys., Saratov State University,
Astrakhanskaya ul., 83, Saratov\, RUSSIA.}\,
Internet address: dvoeglazov@main1.jinr.dubna.su
}

\normalsize{

}

\end{document}